\documentclass[prd,aps,floats,twocolumn]{revtex4}
\usepackage{latexsym}
\usepackage{amssymb}
\usepackage[dvips]{graphicx}
\begin{document}
\newcommand{\beq}{\begin{equation}}
\newcommand{\eeq}{\end{equation}}
\newcommand{\ie}{{\sl i.e\/}}
\newcommand{\half}{\frac 1 2}
\newcommand{\lag}{\cal L}
\newcommand{\ove}{\overline}
\newcommand{\et}{{\em et al}}
\newcommand{\Prd}{Phys. Rev D}
\newcommand{\Prl}{Phys. Rev. Lett.}
\newcommand{\Plb}{Phys. Lett. B}
\newcommand{\Cqg}{Class. Quantum Grav.}
\newcommand{\Grg}{Grav....}
\newcommand{\Np}{Nuc. Phys.}
\newcommand{\Fp}{Found. Phys.}
\renewcommand{\baselinestretch}{1.2}

%\hfill \emph{filename:darklight3.tex} \hfill{\textbf{10/10/06}}
%@twocolumnfalse\endcsnam
\title{Cosmological Effects of Nonlinear Electrodynamics}
%and the acceleration of the universe}

\author{M. Novello, E. Goulart, J. M. Salim}
\affiliation{Instituto de Cosmologia Relatividade Astrofisica (ICRA-Brasil/CBPF) \\
 Rua Dr. Xavier Sigaud, 150, CEP 22290-180, Rio de Janeiro, Brazil}
\author{S. E. Perez Bergliaffa}
\affiliation{Departamento de Fisica Teorica, Universidade do Estado do Rio de Janeiro,
R. S\~ao Francisco Xavier, 524,
Maracan\~a - CEP: 20559-900 - Rio de Janeiro,  Brazil.}

\vspace{.5cm}

%\small{\date{\today}}
%\maketitle

\begin{abstract}
It will be shown that a given realization of nonlinear
electrodynamics, used as source of Einstein's equations,
generates a cosmological model with interesting features, namely a
phase of current cosmic acceleration, and the absence of an
initial singularity, thus pointing to a way to solve two important
problems in cosmology.
\end{abstract}
%\vskip2pc]
%\date{\today}

\vskip2pc
 \maketitle

\section{Introduction}

In spite of the success of the
Standard Cosmological Model (SCM), several problems remain to be solved, most notably the
cause of the current acceleration stage, and the initial singularity.
The observation that the universe is undergoing a phase
of accelerated expansion has sparkled an intense activity directed
to investigate the possible candidates to fuel the
acceleration. One possibility is the
modification of the gravitational action by the addition to the Einstein-Hilbert
action of terms that depend on some power of the curvature. The
simplest of these models \cite{carroll1} with action given by
$$
S =\frac{M_{\rm Pl}^2}{2} \int \sqrt{-g} \left( R -
\frac{\alpha ^4}{ R} \right) d^4x,
$$
has been shown to disagree with solar system observations
\cite{dolgov}. Later it was proved that more general models
\cite{carroll2,Das}, such as those given by the Lagrangian
\begin{equation}
{\cal L} = \sum_n c_{n} \, R^{n}, \label{4jan1}
\end{equation}
can describe a phase of accelerated expansion (controlled by the
negative powers of $n$) and produce other modifications in the
strong regime (where terms with $n > 0$ are dominant).

A second possibility relates the acceleration to the matter sector
of the theory, either by incorporating a cosmological constant or by
postulating the existence of matter fields with peculiar properties.
In the latter case, the most popular choice is a scalar field in the
presence of a potential \cite{ratra}, or with a nontrivial
Lagrangian (see \cite{martin} and references therein). Yet another
possibility is a vector field, as discussed in \cite{picon}.
Although the addition of these fields to the action yields
acceleration (and in some cases dark matter as a
byproduct \cite{martin}), the need of (yet to be observed) matter
with unusual properties is certainly a hindrance.

Another problem in the description furnished by the SCM is that of
the initial singularity. It is a well-known fact that under certain
assumptions, the SCM unavoidably leads to a singular behaviour of
the curvature invariants in what has been termed the Big Bang. This
is a highly distressing state of affairs, because the presence of a
singularity entails the breakdown of all known physical theories. There
are hopes that quantum gravity will solve this deficiency but, since
this theory is still incomplete, one is obliged to explore
alternative routes in order to avoid the initial singularity.

We shall show in this article that nonlinear electrodynamics
(NLED) can be useful in the discussion of
possible solutions to these two problems of the SCM.
Regarding the accelerated expansion,
an alternative to the two possibilities presented
above was given in the model introduced
in \cite{nosso}, where the action for the electromagnetic field was
modified by the addition of a new term, namely
\beq S = \int
\sqrt{-g} \left( - \frac F 4+ \frac \gamma F \right) d^4x,
\label{action}
\eeq with $ F \equiv F_{\mu\nu} \, F^{\mu\nu}.$
Through a suitable averaging process, this action yields an
accelerated expansion phase for the evolution of the universe in the
weak-EM-field regime. It also correctly describes the electric field
of an isolated charge \cite{nosso} without the need of averaging.

Scenarios that avoid the initial singularity have been intensely
studied over the years. As examples of the latest realizations we
can quote the pre-big-bang universe \cite{pbb} and the ekpyrotic
universe \cite{ekpy}.  While these models are based on deep
changes to known physics (involving new entities as scalar fields
or branes) the model we present here relies instead on an unique
entity, the electromagnetic field. The new ingredient that we
introduce is a modification of the dynamics, which differs from
that of Maxwell in certain regimes. Specifically, the Lagrangian
we will work with is given by
\beq {\cal L} =  \alpha \, F^{2}
-\frac{1}{4} \, F + \frac{\gamma}{F}. \label{lag}
\eeq
The
dimensional constants $\alpha$ and $\gamma$ are to be determined
by observation. We shall see that in
Friedmann-Lem\^aitre-Robertson-Walker (FLRW) geometry the first
term dominates in very early epochs, the Maxwell term dominates in
the radiation era, and the last term is responsible for the
accelerated phase.

It will be shown in this article that the Lagrangian given in Eqn.(\ref{lag})
yields a unified
scenario to describe both the acceleration
of the universe (for weak fields, as discussed above) and the
avoidance of the initial singularity as a consequence of its
properties in the strong-field regime (see also \cite{vitorio}).
The plan of the article is as follows. In section II we present some
consequences of applying Tolman average procedure to a nonlinear EM field
configuration in a FLRW geometry, we describe its energy content and
show how nonlinearities change the equation of state of the
electromagnetic field. In section III we present the
notion of magnetic universe, analyze the related
energy density and pressure and show the
unexpected result that the dependence of the magnetic field on the
scale factor $a(t)$ is the same irrespectively of the form of the
Lagrangian. Section IV deals with the conditions of the existence of
a bouncing and the acceleration for a generic fluid. In Section V
we present a simple model of a bouncing universe dominated by the
nonlinear properties of the averaged magnetic field, as well as
a discussion about a cyclic universe and its corresponding
properties in this model. Section VI presents the modifications that
nonlinearities of the electromagnetic field in the weak regime
produce for a late phase of the universe. Several interesting topics
related to nonlinear electromagnetism (such as the field of a point charge and causality)
are discussed in the appendices.

%Indeed, an old universe
%is dominated by powers lower than one on the invariant $F.$ This is
%a direct consequence of the fact that a magnetic universe yields a
%unique and same dependence of the field on the scale factor
%independently of its corresponding dynamics, as we will show. This
%means that weak fields can be of fundamental importance on the
%dynamics of the scale factor for latter epochs.
Section VII presents
a complete scenario in which both problems - bouncing and
acceleration - are treated in a single and an
unified theory.

\section{The average
procedure and the fluid representation}
\label{ave}
The effects of a nonlinear electromagnetic theory in a
cosmological setting have been studied in several articles
\cite{portuga1}.
%Some attention has been devoted to the dynamics of a
%universe governed by matter described by Born-Infeld theory
%\cite{portuga1}.
In particular, it has been shown that under certain assumptions (as
for instance the existence of a compactified space) a Born-Infeld
field as a source can model the accelerated expansion of the
universe \cite{portuga1}, although this
is not possible in the original formulation of the theory,
as we will show in Appendix. Notice that due to the
isotropy of the spatial hyper-surfaces
%We shall see next how a given NLEM theory can be source of Eintein's equations.
of FLRW
geometry,
an average procedure is needed if electromagnetic fields
are to act as a source of gravity. Given a generic gauge-independent
Lagrangian ${\cal L} = {\cal L}(F,G)$, written in terms of the two invariants $F
\equiv F_{\mu\nu} F^{\mu\nu}$ and $G \equiv F^{\mu\nu}
F^{*}_{\mu\nu}$, the associated energy-momentum tensor, defined
by
\begin{equation}
T_{\mu\nu} = \frac{2}{\sqrt {-g}} \frac{\delta {\cal L}
\sqrt{-g}}{\delta g^{\mu\nu}}, \label{n10}
\label{NL2}
\end{equation}
reduces to
\begin{equation}
T_{\mu\nu} = -4 \, {\cal L}_{F} \, F_{\mu}{}^{\alpha} \, F_{\alpha\nu} +
(G\, {\cal L}_{G} - {\cal L} ) \, g_{\mu\nu}. \label{NL3}\end{equation}

Following a standard procedure \cite{robertson} we define the
volumetric spatial average of a quantity $X$ at the time
$t$ by
\beq \overline X \equiv \lim_{V\rightarrow V_0} \frac 1 V
\int X \sqrt{-g}\;d^3x,
\eeq where $V = \int \sqrt{-g}\;d^3x$ and
$V_0$ is a sufficiently large time-dependent three-volume. In this
notation, for the electromagnetic field to act as a source for the
FLRW model we need to impose that
\beq \overline E_i =0, \;\;\;
\overline H_i =0,\;\;\; \overline{E_i H_j}=0, \eeq \beq
\overline{E_iE_j}=-\frac 1 3 E^2 g_{ij}, \;\;\;\overline{H_iH_j} =
-\frac 1 3 H^2 g_{ij}.
\eeq
With these conditions, the
energy-momentum tensor of the EM field associated to  ${\cal L} = {\cal L}(F,G)$
can be written as that of a perfect fluid,
\begin{equation}
T_{\mu\nu} = (\rho + p) v_\mu v_\nu - p\; g_{\mu\nu}, \label{NL33}
\end{equation}
 where
\begin{eqnarray}
\rho &=& - {\cal L} + G\,{\cal L}_{G} - 4 {\cal L}_{F} E^{2},  \\
p &=& {\cal L} - G\,{\cal L}_{G} - \frac{4}{3} \, ( 2 H^{2} - E^{2}) \, {\cal L}_{F},
\label{NL4}
\end{eqnarray}
and
${\cal L}_A\equiv \frac{d{\cal L}}{dA}$, $A=F,G$.
%The
%authors of \cite{vitorio} showed that nonlinear corrections coming
%from an Euler-Heisenberg type of Lagrangian can be important in
%the very early universe, leading to the avoidance of the
%singularity.
In the present work we shall restrict to the case of a nonlinear
theories defined by ${\cal L}={\cal L}(F).$ In such a case, the
energy-momentum is diagonal, with
\beq \rho = -{\cal L}-4E^2{\cal L}_F,\;\;\;\;p = {\cal
L}+\frac 4 3 (E^2-2H^2){\cal L}_F \label{emse} .\eeq

\section{Magnetic universe}
\label{mag}
%A large class of theories can be written in terms of finite
%polynomial of the invariants. Let us limit our analysis here to a
%sub-class of them  defined by the Lagrangian \beq {\cal L} =
%\sum_k c_k F^k, \label{action2} \eeq where the sum can have both
%positive and negative powers.

%Such possibility of allowing for negative power values is not in
%the main stream of suggestions of nonlinear theories but is
%imposed here for the need to generate negative contribution for
%the pressure ... compare com curvatura.

A particularly interesting case of the scenario outlined in the previous section
occurs when only the average of the magnetic part is different from
zero. We shall call such a case a \emph{magnetic universe}. This
situation turns out to be relevant in cosmology, because the
electric field is screened by the charged primordial plasma, while
the magnetic field lines are frozen \cite{prim}. In spite of this
fact, we shall devote some attention to the mathematically
interesting case in which $E^2 = \sigma^{2} H^{2} \neq 0$ in
Appendix 1.
%, among which the
%dependence on $\sigma^{2}$ of the equation of state - relating the
%density of energy with the pressure -- except in the very
%particular case of Maxwell linear electrodynamics.

%The magnetic universe has some nice properties
Before discussing the magnetic universe, we shall show that
when the dynamics of the EM field is given by the following series
\begin{equation}
{\cal L} = \sum_k \, c_{k} \, F^{k}, \label{19dez1}
\end{equation}
(where $k$ takes values on the integers)
the fluid can be interpreted as composed of $k$
non-interacting fluids. In order to show that
this property is indeed valid, we shall work with
the standard form of the FLRW geometry in Gaussian coordinates
\begin{equation}
ds^{2} = dt^{2} - a(t)^{2} \, \left( dr^{2} + \chi^{2} d\Omega^{2}
\right). \label{dez181}
\end{equation}
%The expansion factor $\theta$
%reduces, in the present case, to the derivative of the logarithm of
%the scale factor:
%\begin{equation}
%\theta \equiv v^{\mu}_{\;;\mu} = 3 \, \frac{\dot{a}}{a}.
%$\label{dez182}
%$\end{equation}
The conservation of the energy-momentum tensor projected in the
direction of the co-moving velocity $v^{\mu} = \delta^{\mu}_{0}$
yields
\begin{equation}
\dot{\rho} + (\rho + p) \theta = 0 \label{M1},
\end{equation}
where
$\theta \equiv v^{\mu}_{\;;\mu} = 3 \, \frac{\dot{a}}{a}$.

Using the values of energy density and pressure given in
Eqn.(\ref{emse}) along with Eqn.(\ref{19dez1})
we obtain that
\beq
\rho = \sum_k \rho_{k},\;\;\;\;\;\;\;
p = \sum_k p_{k}, \label{19dez2}
\eeq
where, for the magnetic universe, we have
\begin{eqnarray}
\rho_{k} &=& -c_{k} 2^{k} H^{2k} ,\nonumber \\
p_{k} &=& c_{k}  \, 2^{k} H^{2k} \left(1 - \frac{4k}{3} \right).
\label{19dez3}
\end{eqnarray}
From the conservation equation (\ref{M1}) we obtain
\begin{equation}
{\cal L}_{F} \left[(H^{2})\dot{} +  4 \, H^{2} \,
\frac{\dot{a}}{a}\right] = 0.
\label{M2}
\end{equation}
The important result that follows from this equation
is that the dependence on the specific form
of the Lagrangian appears as a multiplicative factor.
This property allows us to obtain the dependence
of the field with the scale factor independently of the particular
form of the Lagrangian, since Eqn.(\ref{M2}) yields
\begin{equation}
 H = \frac{H_{0}}{a^{2}}.
\label{dez183}
\end{equation}
This property implies that it is possible to associate to each power
$k$ an independent fluid with energy density $\rho_{k}$  and
pressure $p_{k}$ in such a way that the corresponding equation of
state, which is preserved throughout the expansion of the universe,
is given by \beq p_k = \left( \frac{4k}{3} - 1 \right) \rho_k.
\label{partialp} \eeq
%The modifications of the behavior of
%the electromagnetic field contained in theories described by
%powers of $F$ that allow for such property are precisely those
%that we named \textbf{dark light. }
Before proceeding to examine the consequences of these fluids in the
expansion of the universe, let us turn to the analysis of the
conditions needed for a bounce and for an accelerated expansion.

\section{Conditions for bouncing and acceleration}

We shall begin by setting the conditions needed to have acceleration.
From Einstein's equations, the acceleration of the FRWL universe is
related to its matter content by
\beq 3 \frac{\ddot a}{a} = -
\half (\rho + 3 p). \label{acc} \eeq
Thus, in order to have an
accelerated expansion, the source must satisfy the constraint
$(\rho + 3 p)<0$. In terms of the quantities defined in
Eqn.(\ref{emse}),
%\begin{equation}
%\rho + 3 p = 2(L-4H^2L_F). \label{TA1}
%\end{equation}
% Hence the
this
constraint
%$(\rho + 3 p)<0$
translates into \beq {\cal L}_F > \frac{{\cal L}}{4H^2}. \eeq Any
nonlinear electromagnetic theory that admits a configuration
satisfying this inequality will yield accelerated expansion.
Referring now to the particular Lagrangian given in
Eqn.(\ref{19dez1}), Eqn.(\ref{partialp}) shows that if the
$k$-component with $k < \frac{1}{2}$ dominates the evolution of the
universe, there will be accelerated expansion (assuming that
$\rho_k>0$).

The conditions to have a bouncing universe
follow from Raychaudhuri's equation, which states that the expansion
factor must obey the equation
\begin{equation}
\dot{\theta} + \frac{1}{3} \, \theta^{2} = \half (\rho + 3 p).
\label{4jan2}
\end{equation}
In addition to this equation, the energy content of the fluid must satisfy
the constraint
\begin{equation}
\rho =  \frac{1}{3} \, \theta^{2} + \frac{3 \epsilon}{a^{2}},
\label{19junho1}
\end{equation}
which comes from the $0-0$ component of Einstein equation in the
FLRW framework. Thus two conditions must be fulfilled
for the bounce to be a minimum of the
scale factor $
(\rho + 3 p )_{b}$ and $\rho_{b} = \frac{3\epsilon}{a_{b}^{2}}$
\footnote{We assume here that the cosmological constant is zero. See \cite{ademir}
for the analysis with $\Lambda\neq 0$.}.

Having established the conditions for accelerated expansion and for a bounce,
we shall show next that
the magnetic
universe with the Lagrangian given in
Eqn.(\ref{lag}) satisfies both conditions.

\section{A bouncing universe}
\label{bounce}
Let us briefly review the scenario proposed in
\cite{vitorio}, where a nonsingular universe was presented,
based on the Lagrangian
\begin{equation}
L = -\frac{1}{4}\,F + \alpha\,F^2,   \label{Order2}\end{equation}
where $\alpha$ has the dimension of (length)$^4$.
Notice that
%The dimensional constant $\alpha$ is
%to be determined by the observation.
Maxwell's electrodynamics can be obtained from Eqn.(\ref{Order2}) by
setting $\alpha=0$, or recovered in the limit of small fields. Using
the average process introduced in Sect.(\ref{ave}),
it follows that
%tum tensor
%for nonlinear electromagnetic theories %\cite{Novello}
%reads
%\begin{equation}
%\protect\label{Tmunu} T_{\mu\nu}=-4\,L_F\,F_\mu\mbox{}^\alpha
%F_{\alpha\nu} + (GL_G-L)\,g_{\mu\nu},
%\end{equation}
%The homogeneous Lagrangian (\ref{Order2}) requires some spatial
%averages over large scales, as we did before. If one intends to
%make similar calculations on smaller scales then either more
%involved non homogeneous Lagrangians should be used
%or some additional magnetohydrodynamical effect \cite{Thompson} %,Subramanian}
%should be devised in order to achieve correlation \cite{Jedamzik}
%at the desired scale. Since the average procedure is independent
%of the equations of the electromagnetic field we can use the above
%formulae (\ref{meanEH})--(\ref{meanH2}) to arrive at a counterpart
%of expression (\ref{Pfluid}) for the non-Maxwellian case. The
%average energy-momentum tensor is identified as a perfect fluid
%(\ref{Pfluid}) with modified expressions for the energy density
%$\rho_\gamma$ and pressure $p_\gamma$ as we did arriving at the
%expressions
\begin{eqnarray}
\rho &=& \frac{1}{2} \, H^2 \,(1 - 8\,\alpha\,H^2),
\label{rho}\\[1ex]
\protect\label{P} p &=& \frac{1}{6} \,H^2 \,(1 -
40\,\alpha\,H^2).
\end{eqnarray}
It was shown in Sect.(\ref{mag})
that $ H=H_0/a^2,$ where $H_0$ is a constant.
%Equation (\ref{H->A})
%This relation holds as well in the Maxwell case.
Inserting this result into Friedmann's equation
through Eqn.(\ref{rho})
leads to
\begin{equation}
 \dot{a}^2=\frac{H_0^2}{6\,a^2} \left(1-\frac{8\alpha
H_0^2}{a^4}\right)-\epsilon.
\label{eqA2}
\end{equation}
Since the right-hand side of this equation cannot be
negative it follows that, regardless of the value of $\epsilon$,
for $\alpha>0$ the scale-factor $a(t)$ cannot be arbitrarily
small. The solution of Eqn.(\ref{eqA2}) is implicitly given as
\begin{equation}
\label{solution} t=\pm\int_{a_0}^{a(t)}\frac{\textstyle dz}
{\sqrt{\textstyle\frac{\textstyle H_0^2}{\textstyle6z^2}
-\frac{\textstyle8\alpha H_0^4}{\textstyle6z^6}-\epsilon}},
\end{equation}
where $a(0)=a_0$.
%\texttt{The linear case (\ref{A(t)Maxwell}) can be
%achieved from Eq.\ (\ref{solution}) by setting $\alpha=0$.
%which yields $A(t)=A_o\sqrt{(1+\epsilon^2)t/t_o-\epsilon(t-t_o)^2}$.
For the Euclidean section, Eqn.(\ref{solution}) can be easily solved
to yield
\begin{equation}
\label{A(t)} a^2 = H_{0} \,\sqrt{\frac{2}{3}\,t^2 + 8\,\alpha}.
\end{equation}
From Eqn.(\ref{dez183}), the average strength of the magnetic
field $H$ evolves with time as
\begin{equation}
\protect\label{H(t)} H^2 = \frac{3}{2}\,\frac{1}{{t^2} +
12\,\alpha}.
\end{equation}
Expression (\ref{A(t)}) is singular for $\alpha<0$, since there
exist a time $t=\sqrt{-12\alpha}$ for which $a(t)$ is arbitrarily
small. For $\alpha>0$ the radius of the universe attains a minimum
value at $t=0$, given by
\begin{equation}
\label{Amin} a^2_{\rm min} = H_{0} \, \sqrt{8\,\alpha}.
\end{equation}
Therefore, the actual value of $a_{\rm min}$ depends on $H_0$, which
turns out to be the sole free parameter of this model.
The energy density $\rho$ given by Eqn.(\ref{rho}) reaches its
maximum value $\rho_{\rm max}=1/64\alpha$
%\begin{equation}
%\protect\label{maxRho}
%\rho_{max} = \frac{1}{16 \,\mu}
%\end{equation}
at the instant $t=t_0$, where
\begin{equation}
\label{tc} t_{0} = \,\sqrt{12\,\alpha}.
\end{equation}
For smaller values of $t$, the energy density decreases, vanishing
at $t = 0$, while the pressure becomes negative.
%As remarked before,
Only for values of $t$ such that
$t\,\raisebox{-0.5ex}{$\stackrel{<}{\scriptsize\sim}$}
\,\sqrt{4\,\alpha}$, the nonlinear effects are relevant for the
solution. Notice that the solution given in Eqn.(\ref{A(t)}) reduces
to that corresponding to Maxwell's Lagrangian, namely the scale factor for
the radiation era,
for large $t$.

Next we shall show that the Lagrangian given in Eqn.(\ref{Order2})
yields a cyclic universe.
The qualitative properties of the evolution of the scale
factor can be obtained by
interpreting
Eqn.(\ref{eqA2}) as describing
the the motion of a particle with constant energy
$-\epsilon$ and
position $a(t)$ under the influence of the
potential $V$ given by
\begin{equation}
V = - \frac{H_{0}^{2}}{6 a^{2}} \left( 1 - \frac{8 \alpha
H_{0}^{2}}{a^{4}} \right) .\label{19junho3}
\end{equation}
%Equation (\ref{19junho1}) yields the conservation of "energy" $T + V =$
%constant, where $T \equiv \dot{a}^{2}$ and the constant is to be
%identified with the 3-d curvature $\epsilon.$
It follows from the analysis of $V$ that the scale factor $a(t)$ bounces between its minimum
$a_{b}(t)$ and its maximum $a_{B}(t)$ which are provided by the two
real solutions of the cubic equation
\begin{equation}
\epsilon \, x^{3} - \frac{H_{o}^{2}}{6} \, x^{2} + \frac{4}{3} \,
\alpha \, H_{0}^{4} = 0, \label{19junho5}
\end{equation}
where $ x \equiv a^{2}.$

\section{An accelerated Universe}
\label{accel}
As discussed in the previous section, a positive power of
the invariant $F$ in the Lagrangian yields a nonsingular
FLRW cosmology.
In fact, were higher
values of $k$ present, they would dominate the early evolution. Let
us see what happens when there are negative powers of $F$ in the
Lagrangian by analyzing the example given in Eqn.(\ref{lag})
\cite{nosso}. We restrict our analysis to the magnetic universe
and set $\ove{E^2} = 0$, with a residual magnetic field, for the
reasons already explained. In this case,
$$
\rho = \frac{H^2}{2} + \frac{\mu^8}{2} \frac{1}{ BH^2},
$$
where we have set $\gamma \equiv - \mu^8$ \cite{gamma}.
%From the conservation law
%$$
%\dot \rho + 3(\rho + p ) \frac{\dot a}{a} = 0,
%$$
Since $H = H_0 / a^2$, the evolution of the density
with the scale factor is
\beq
\rho = \frac{H_0^2}{2}\;
\frac{1}{a^4} + \frac{\mu^8}{2H_0^2}\; a^4. \label{density}
\eeq
For small $a$ it is the ordinary radiation term that dominates.
The $1/F$ term takes over only after $a=\sqrt{H_0}/\mu$, and grows
without bound afterwards. In fact, the curvature scalar is
$$
R = T^\mu_{\;\mu} = \rho-3p = \frac{4\mu^8}{H_0^2}\; a^4.
$$
Using the density given in Eqn.(\ref{density}) in Eqn.(\ref{acc}) gives
$$
3 \frac{\ddot a}{a} +  \frac{H_0^2}{2}\; \frac{1}{a^4} - \frac 3 2
\frac{{\mu^8}}{H_0^2}\; a^4=0.
$$
To get a regime of accelerated expansion, we must have
$$
 \frac{H_0^2}{a^4} -  3
\frac{{\mu^8}}{H_0^2}\; a^4 < 0,
$$
which implies that the universe will accelerate for $a>a_c$, with
$$
a_c = \left(\frac{H_0^4}{3\mu^8}\right)^{1/8}.
$$
The examples of positive and negative powers of $F$
analyzed in the two previous sections
suggest
that it is worth analyzing
a complete scenario constructed by
combining these cases in a single phenomenological scenario.
This will be done in the next section.

\section{A complete scenario}

\begin{figure}[h]
\begin{center}
\includegraphics[angle=-90,width=0.5\textwidth]{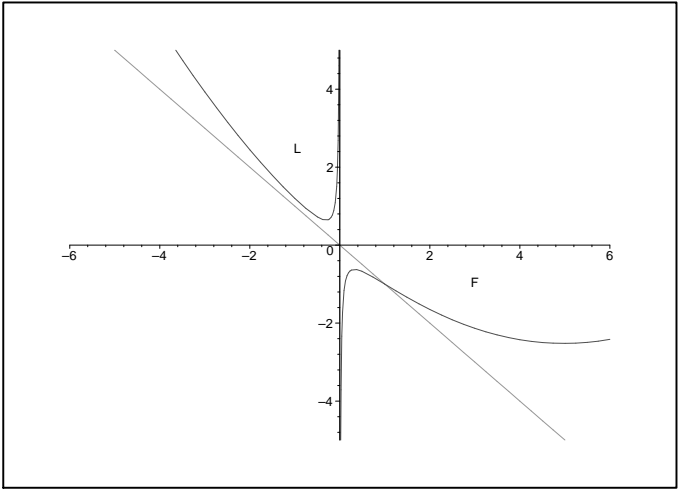}
\caption{The Lagrangian given in Eqn.(\ref{11jan1}) as a function of
$F$.}
\label{lagr}
\end{center}
\end{figure}

%There is no doubt that electromagnetic radiation described by a
%maxwellian distribution has driven the cosmic geometry for a
%period.
We shall analyze here the evolution of the toy model
generated by the Lagrangian
\begin{equation}
{\cal L} = -\frac{1}{4}\,F + \alpha\,F^2 - \frac{\mu^{8}}{F}
\label{11jan1}
\end{equation}
with the dependence of the magnetic field on the scale factor
given by $H = H_0 / a^2$. In
Fig.(\ref{lagr}) we plot the Lagrangian as a function of the value of the
field.
The derivative $dL/dF$ has three zeros in points $a, b,c.$
In these points $\rho + p$ vanishes.  In the case of pure magnetic
universe the value of $F$ is always positive.

As shown in Appendix III, the theory described by Eqn.(\ref{11jan1})
applied to the case of a point charge gives results that are in agreement
with observation. In the cosmological case, it
implies the existence
of three distinct epochs according to the value of the field.
In the bouncing era
the value of the curvature scalar is small and the
volume of the universe attains its minimum. The energy density
and the pressure are dominated by the terms coming from the
quadratic lagrangian $F^{2}$ and are approximately given by
\begin{eqnarray*}
\rho & \approx & \frac{H^{2}}{2} \, ( 1 - 8 \alpha^{2}\, H^{2}),
\nonumber \\
p & \approx & \frac{H^{2}}{6} \, ( 1 - 40 \alpha^{2}\, H^{2}).
%\label{2Maio11}
\end{eqnarray*}
In the radiative era,
the standard Maxwellian term dominates.
Due to the dependence on $a^{-2}$ of the field, this phase
is defined by $H^{2} >> H^{4}$ yielding the approximation
%\begin{eqnarray}
$$
\rho  \approx  \frac{H^{2}}{2},
% \nonumber \\
\;\;\;\;\;\;\;\;\;\;\;
p  \approx  \frac{H^{2}}{6}.
%\, \label{3Maio11}
%\end{eqnarray}
$$
Finally, the
accelerating era
takes place when the universe becomes large, and the $1/F$ term
dominates. In such a case,
%\begin{eqnarray}
$$
\rho  \approx  \frac{1}{2} \, \frac{\mu^{8}}{H^{2}},
%\nonumber \\
\;\;\;\;\;\;\;\;\;\;
p  \approx - \frac{7}{6} \, \frac{\mu^{8}}{H^{2}}.
%\label{4Maio11}
%\end{eqnarray}
$$
%In fig. 2 we plot the density of energy as function of the scalar
%of curvature exhibiting the different phases of transition. The
%time dependence of $a(t)$
We can analyze the whole evolution using the effective potential,
given by
%\subsection{Potential}
%It will be more direct to examine the effects of the magnetic
%universe controlled by the above lagrangian if we undertake a
%qualitative analysis using an analogy with classical mechanics.
%Friedmann's equation reduces to the set
%Note that Friedmann's equation can be written as \beq \dot
%a^2+V(a) = -\epsilon, \eeq where
\beq
V(a)=\frac{A}{a^6}-\frac{B}{a^2}-Ca^6 \label{pot}. \eeq
The constants in $V(a)$ are given by
$$
A=4\alpha H_0^4,\;\;\;\;\;\;
B=\frac{1}{6}H_0^2,\;\;\;\;\;\;C=\frac{\mu^8}{2H_0^4},
$$
and are all positive.

\begin{figure}[h]
\begin{center}
\includegraphics[angle=-90,width=0.5\textwidth]{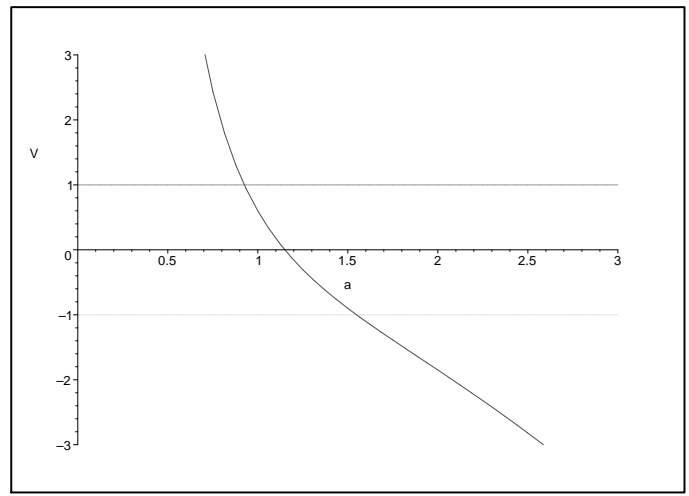}
\caption{Plot of the effective potential for
$D_V>0$, $D_{V'}>0$.}
\label{potencial2}
\end{center}
\end{figure}

\begin{figure}[h]
\begin{center}
\includegraphics[angle=-90,width=0.5\textwidth]{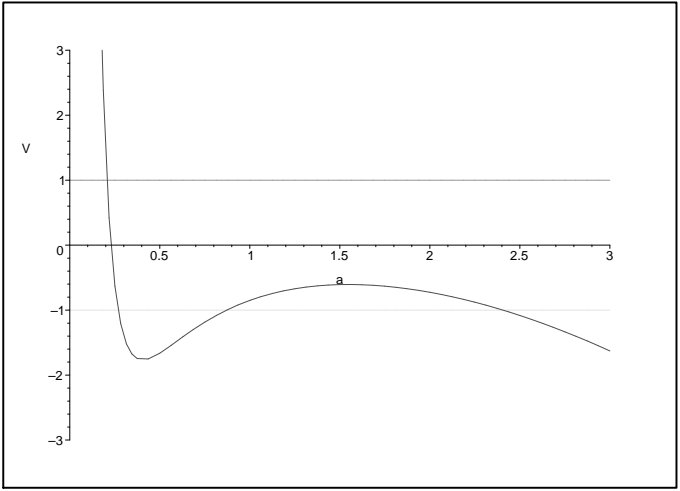}
\caption{Plot of the effective potential for $D_V>0$, $D_{V'}=0$.}
\label{potencials}
\end{center}
\end{figure}

The analysis of $V(a)$ and its derivatives implies
solving polynomial equations in $a$, which can be reduced to
cubic equations through the substitution $z=a^4$.
The existence and features of the roots of such
equations are discussed in \cite{bir}.
A key point to the analysis is the sign of $D$, defined as follows.
For a general cubic equation
$$x^3+px=q,
$$
the discriminant $D$ is given by
$$
D=\left(\frac p 3 \right)^3+\left(\frac q 2 \right)^2.
$$
We will denote by $D_V$ the discriminant
corresponding to the potential and $D_{V'}$ that of the derivative
of V. From the behaviour of the potential and its derivatives for
$a\rightarrow 0$ and $a\rightarrow \infty$ we see that only one or
three zeros of the potential are allowed. In the case of one zero
(described by $D_V>0$, $D_{V'}>0$), there must be an inflection
point, and the qualitative plot of the potential is
given in Fig. \ref{potencial2}.
The position of the inflection point depends on the values of the parameters of the model.
We see that this model is nonsingular for any value of $\epsilon$,
but it does not display a transition from an
accelerated regime to a non-accelerated one, so we shall move on to the other possibility
(given by $D_V>0$, $D_{V'}=0$),
displayed in Fig. \ref{potencials}.
This figure shows the qualitative behavior of the potential for
typical values of the parameters. Again, the model is nonsingular
for any value of $\epsilon$. The maximum and the minimum of $V$ are
always well above the line $\epsilon=1$ so there is no transition
in the acceleration for this value of $\epsilon$. For
$\epsilon=0$ or 1, the model displays a ''coasting period'' about the
transition point.
%, which corresponds to a too-early-time in the
%cosmological evolution for typical values of the parameters.

\section{Causality and NLED}

Before going to the concluding remarks, we would like to
call the attention of the reader to a salient feature of nonlinear electrodynamics in
a cosmological setting.
Most of our description of the universe is based on the behavior
of light in a gravitational field. In a FLRW scenario the
existence of a horizon inhibits the exchange of
information between arbitrary parts of the universe.
The observation of the high degree of isotropy of the CMBR
generated a conflict with causality: different parts of the universe
in the standard FLRW geometry could not have enough time to
homogeneize.
The inflationary scenario was engineered precisely
to solve this problem. But it is worth to mention that
nonlinear theories of
electrodynamics also introduce
a new look into causality, which we will overview very briefly next.
%
%The main lesson we can extract from the analysis of the
%propagation of the wave front in nonlinear theory is contained in
%a theorem that can be demonstrated using Hadamard method to deal
%with the discontinuity of the field and which can be synthesized
%in a single sentence dealing with the modification of the geometry
%generating an effective metric that controls the properties of the
%space. In order to show this let us make a very short resume of
%it.

%\subsubsection{The Effective Metric}
%\label{intro}
%
%Historically, the first example of the idea of effective metric
%was presented in 1923 by W. Gordon. In modern language, the wave
%equation for the propagation of light in a moving non-dispersive
%medium, with slowly varying refractive index $n$ and 4-velocity
%$u^\mu$:
%$$
%\left[ \partial_\alpha \partial^\alpha + (n^2-1) (u^\alpha
%\partial_\alpha)^2 \right] F_{\mu\nu} = 0.
%$$
%Taking the geometrical optics limit, the Hamilton-Jacobi equation
%for light rays can be written as  $ g^{\mu\nu}k_\mu k_\nu = 0 $
%(see \cite{lp} for details), where \beq g^{\mu\nu}= \eta^{\mu\nu}+
%(n^2 - 1) u^\mu u^\nu \label{gordonm} \eeq is the effective metric
%for this problem. It must be noted that only photons in the
%geometric optics approximation move on geodesics of $g^{\mu\nu}$:
%the particles that compose the fluid couple instead to the
%background Minkowskian metric.
Let us take as an example the action \footnote{We could have
considered $L = L(F,G)$ instead, where $G \equiv F^{*}_{\mu\nu}
F^{\mu\nu}$. This case is studied in detail in \cite{vitorio}.}
\beq S = \int
\sqrt{-\gamma}\; L(F) \;d^{4}x \protect \label{N1}
\end{equation}
($\gamma$ is the determinant of the background metric).
%Varying this action w.r.t. the
%potential $W_{\mu}$,
% related to the field by the expression
%$$
%F_{\mu\nu} = W_{\mu;\nu} - W_{\nu;\mu} = W_{\mu,\nu} -
%W_{\nu,\mu},
%$$
Using either the traditional perturbation method or the method of the
discontinuities devised by Hadamard (see \cite{vitorio} and references therein),
it can be shown that
nonlinear photons do not move on the light cone of the background metric.
Instead they move along the null surfaces of an effective metric given by
\begin{equation}
g^{\mu\nu}= L_{F} \, \gamma^{\mu\nu} - 4 \, L_{FF}\, F^{\mu\alpha}
F_{\alpha}{}^{\nu}. \protect\label{N20}
\end{equation}
Notice that this effective metric is generated solely by the self-interaction of the
electromagnetic field since the metric coincides with the
background geometry when the theory is linear.

For Lagrangians that depend also of
$F^*$, an analogous effective geometry appears, although there may
be some special cases in which the propagation coincides with that
in dictated by the background metric \cite{vitorio}. Another feature of the more general
case $L=L(F,F^*)$ is that birefringence is present. That is, the
two polarization states of the photon propagate in a different
way. In some special cases, there is also bimetricity (one
effective metric for each state). In some special cases of
nonlinear theories (such as Born-Infeld electrodynamics) a single
metric can appear \cite{vitorio}.
%\subsection{Causal properties in the fundamental state}

Let us apply the effective metric to
%for the causal structure in
the case in which the electromagnetic field
rests on its fundamental state.
%From the calculation made in
%previous  chapter the photons propagate in an effective geometry
%which is given by equation (\ref{N20}).
In the case of Lagrangian
\beq
{\cal L} = -\frac{1}{4}\,F - \frac{\mu^{8}}{F},
\label{lnl}
\eeq
the fundamental state is the particular solution
$$ F^{2} = 4 \mu^{8},  $$
which corresponds to an energy-momentum tensor equivalent to a
fluid distribution characterized by the condition $ \rho + p = 0$
and generates a de Sitter geometry for the background metric
$\gamma_{\mu\nu}$ as seen by all forms of matter and energy content,
as far as we neglect the gravitational influence of such remaining
matter and energy.

For the Lagrangian given above, the effective metric tensor takes the form:
%\begin{equation}
$$
g^{\mu\nu}_{(eff)}= (- \, \frac{1}{4} + \frac{\mu^{8}}{F^{2}}) \,
g^{\mu\nu}
 +  \frac{8 \mu^{8}}{F^{3}} \,  F^{\mu\alpha} F_{\alpha}{}^{\nu},
%\label{CP2}
%\end{equation}
$$
which in the case of the fundamental state reduces to
%\begin{equation}
$$
g^{\mu\nu}_{(eff)} = \pm \frac{1}{\mu^{4}} \,  F^{\mu\alpha}
F_{\alpha}{}^{\nu}.
%\label{CP3}
%\end{equation}
$$
%Hence, the photons do not propagate in a deSitter geometry
%but instead in an effective metric which is provided by the form:
%
This is a very peculiar and interesting situation:
the fundamental state of the
theory described by the inverse symmetric Lagrangian given in Eqn.(\ref{lnl})
generates a
de Sitter universe as seen by all existing matter with one
exception, namely the photons, which follows geodesics in the above
anisotropic geometry $g^{\mu\nu}_{(eff)}.$

\section{Concluding remarks}

\begin{figure}[h]
\begin{center}
\includegraphics[angle=-90,width=0.5\textwidth]{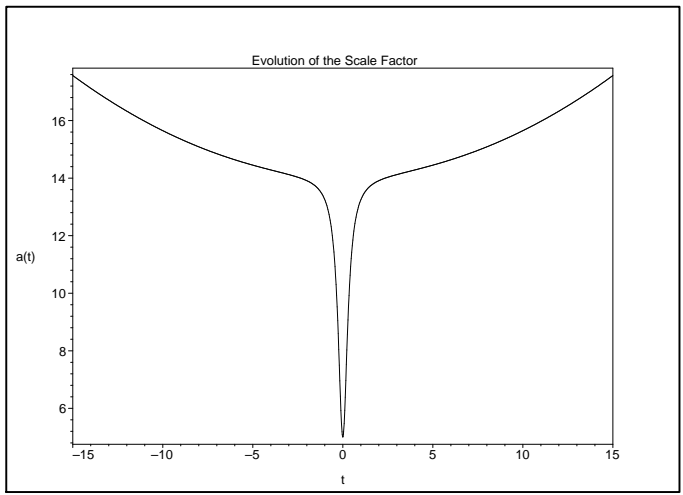}
\caption{Qualitative plot of the scale factor as a function of $t$.}
\label{scale}
\end{center}
\end{figure}

The standard cosmological model furnishes a rather complete picture of our universe, but
it does not solve yet some important problems. Among these, the most notable are the initial singularity,
and the accelerated expansion. We have shown here that a specific realization of nonlinear
electrodynamics yields a unified description of the evolution of the universe, in the sense
that with this single matter source we obtain a qualitative description of three of the main
phases: bounce (instead of singularity), radiation, and acceleration.
The description of the matter content with a single field has the advantage,
when compared to multi-field models, that in principle
less parameters and initial conditions are needed.
The above mentioned features are shown in the plot of the scale factor in terms of
the cosmological time (see Fig.\ref{scale}).

%Furthermore, we have shown that the chosen nonlinear theory is capable of describing
%the static electric field of a point charge in agreement with observation.
Notice that in spite of the fact that the fluid as a whole would have a time-dependent
equation of state, only one component of the fluid dominates in each of the epochs,
leading to an effective equation of state with constant parameter.

We have presented here a toy model which shows that nonlinear electrodynamics could be
used to build a complete scenario of the evolution of the universe. We leave the matter
of the agreement between the values of the equation of state
parameters used here and those that
follow from observation to a future publication. Eventual
modifications of the model by the addition of
new terms in the nonlinear Lagrangian, following the same one-field phenomenological
approach used here, will also be considered, if suggested by observation.

%COLOCAR AQUI TENSOR TMUNU E PROPRIEDADES

\section*{Appendix I: Almost Magnetic Universe (AMU)}
\label{sigma}
Although the main goal of this paper is the study of the
pure
magnetic universe, it is worth to make some comments on the
case in which there is an electric component such that
$$ E^{2} = \sigma^{2} \, H^{2}, $$
where $ 0 < \sigma^{2} < 1.$
%We will do this in the appendix for
%the general case of the series dealt with in equation
%(\ref{19dez1}) where we show that the relationship between the
%density of energy and pressure depends now not only on the power
%$k$ - that characterizes the dependence on $F$ that appears in the
%lagrangian (\ref{19dez1})- but also on the value of the
%coefficient $\sigma^{2}$ that gives the ratio between the electric
%and the magnetic parts of the field. Here we limit to the case
%
%Let us start by the Lagrangian
%$$ {\cal L} = -1/4 \, F + \alpha \, F^{2}.$$ From Eqns.(\ref{NL4})
%we obtain
%\begin{equation}
%\rho = \frac{(\sigma^{2} + 1)}{2} \, H^{2} + 4 \alpha \, H^{4} \,
%(\sigma^{2} - 1) \, (3 \sigma^{2} + 1) \label{6jan2}
%\end{equation}
%\begin{equation}
%p = \frac{(\sigma^{2} + 1)}{6} \, H^{2}  + \frac{4}{3} \alpha \,
%H^{4} \, (\sigma^{2} - 1)( 5 - \sigma^{2}) \label{6jan3}.
%\end{equation}
%Let us remark first that the volumetric enthalpy, given in this case by
%$$
%\varpi = \frac 2 3 H^2(1+\sigma^2)-\frac{32}{3}\alpha H^4 (1-\sigma^4);
%$$
%is minimized by the state with $\sigma = 0$, that is, by the magnetic universe.
%The same thing happens in
%with the Born-Infeld Lagrangian, given by
%\begin{equation}
%{\cal L}_{bi} = - \, \beta^{2} \sqrt{X} +  \beta^{2} \label{6jan6}
%\end{equation}
%where $X = 1 + F/2\beta^{2}$ which yields
%\begin{equation}
% \omega = \frac{2}{3} \,(\sigma^{2} + 1)\,  \frac{H^{2}}{\sqrt{X}} \label{6jan7}
%\end{equation}
%It then follows that the state which corresponds to $\sigma = 0$
%minimizes the enthalpy.
Let us assume that
the Lagrangian is given by the power series
in Eqn.(\ref{19dez1}), and study
the relation between the energy density and the pressure.
It follows from Eqns.(\ref{NL4}) and (\ref{19dez1}) that
\begin{eqnarray}
\rho &=& - \sum_k c_{k} F^{k} \, \left( 1 + \frac{2\sigma^{2}}{1 - \sigma^{2}} \, k \right), \nonumber \\
p &=&  \sum_k c_{k} F^{k} \, \left( 1 + \frac{2}{3} \,
\frac{\sigma^{2} - 2}{1 - \sigma^{2}} \, k  \right). \label{14Maio}
\end{eqnarray}
Setting $ p_{k} = \lambda_{k} \, \rho_{k}$ we obtain
\begin{equation}
\lambda_{k} = \frac{1}{3} \,  \frac{  2 k (\sigma^{2} - 2) - 3
(\sigma^{2} - 1)}{ \sigma^{2} - 1 - 2\sigma^{2}\,
k }. \label{AMU1}
\end{equation}
Thus, the equation of state depends on both the power $k$ and on
the partition $\sigma^{2}$, the dependence on these parameters
being caused by the nonlinearity of the theory. Note that in the
linear case (that is, for $k = 1$) the value of the ratio $p/\rho$
reduces to the $\sigma$-independent value $\frac{1}{3}$, as is
clear by inspection of Eqn.(\ref{AMU1}).
%
%\texttt{Thus, for the evolution of the metric structure
%of the universe, the unique influence of such quantity
%$\sigma^{2}$ concerns the normalization of the value of the
%constant $B_{0}$ of  eqn.(\ref{dez183}).}
%
On the other hand, when
there is ''equipartition'', such that $\sigma^{2} = 1 $ then,
independently of the value of the power $k$ it follows that
$\lambda = \frac{1}{3}$. Note that for $ k < 0$
the equipartition is not possible,
the value of the partition is restricted to the open domain $  0 <
\sigma^{2} < 1.$

In two special cases ({\em i.e.} $ k = 1$ or $ \sigma^{2} = 0)$ the dependence of
the field with $a(t)$ is the same. In the general case a
straightforward calculation reduces the conservation equation,
Eqn.(\ref{M1}) in the case in which $\sigma $ is a constant, to
the form:
\begin{equation}
\sum_k c_{k} \, k \, F^{k} \left[ \left(1 + \frac{2k \,\sigma^{2}}{1 -
\sigma^{2} }\right) \, \frac{\dot{F}}{F} + 4 \, \frac{1 + \sigma^{2}}{1
- \sigma^{2}} \, \frac{\dot{a}}{a} \right] = 0. \label{AMU2}
\end{equation}
It follows that different parts of the fluid
interact, so it is
not
possible to treat the fluid as composed by distinct non
interacting parts. The solution of Eqn.(\ref{AMU2}) provides the
implicit dependence of the field in terms of the scalar of
curvature and the constant $\sigma^{2}.$ Let us see this by an example.
In the case of
\beq
{\cal L} =  - \frac F 4+ \frac \gamma F,
\label{action1}
\eeq
the solution of Eqn.(\ref{AMU2})
is given by
\begin{equation}
F^{\frac{3\sigma^{2} - 1}{\sigma^{2} + 1}} \, \left( F^{2} -
\mu^{8} \right)^{-1} a^{-4} = {\rm constant}. \label{dez18ap1}
\end{equation}
The energy density takes the form
\begin{equation}
\rho = \frac{\sigma^{2} + 1}{2} \, H^{2} - \frac{3\sigma^{2} -
1}{2(\sigma^{2} - 1)^{2}} \, \mu^{8} \, H^{-2}, \label{dez18ap2}
\end{equation}
and for the pressure we find
\begin{equation}
p  = \frac{\sigma^{2} + 1}{6} \, H^{2} + \frac{5\sigma^{2} -
7}{6(\sigma^{2} - 1)^{2}} \, \mu^{8} \, H^{-2} .\label{dez18ap3}
\end{equation}
We see from these expressions that although one could separate the
fluid in two parts by defining $\rho = \rho_{1} + \rho_{2}$ and
$p = p_{1} + p_{2}$ the relations between these quantities
depend on the equipartition constant. Indeed, we have
\begin{equation}
p_{1} = \frac{1}{3} \, \varrho_{1} \label{MN1},
\end{equation}
and
\begin{equation}
p_{2} = \frac{5\sigma^{2} - 7}{3(1 - 3\sigma^{2})} \, \varrho_{2}.
\label{MN2}
\end{equation}
Thus, there is an interaction between the fluids.

\section*{Appendix II: Born-Infeld Electrodynamics}

\label{bi}
Although the final version proposed by Born and Infeld of nonlinear
electrodynamics involves both the invariants $F$ and $G$, we
shall use in this section the simplified form $L=L(F)$.
In this case,
\begin{equation}
L_{BI} = \beta^{2} \left( 1 - \sqrt{X} \right), \label{BI1}
\end{equation}
where
\begin{equation}
X \equiv 1 + \frac{1}{2\beta^{2}} \, F . \label{BI2}
\end{equation}
Following Born and
Infeld, have added a constant term into
the Lagrangian in order to eliminate a sort of cosmological
constant and to set the value of the field to zero at
infinity. We will come back to this
question later on.

Using expressions (\ref{emse})
we obtain
\begin{equation}
\rho =  \frac{\beta^{2}}{\sqrt{X}} \, \left(  1 - \sqrt{X} +
\frac{H^{2}}{\beta^{2}} \right), \label{BI3}
\end{equation}
\begin{equation}
p =  \frac{\beta^{2}}{\sqrt{X}} \, \left(  \sqrt{X} - 1  -
\frac{1}{3} \, \frac{H^{2}}{\beta^{2}} \right). \label{BI4}
\end{equation}
It follows that
\begin{equation}
\rho + 3p = \frac{2 \beta^{2}}{\sqrt{X}} \, \left( \sqrt{X}  - 1
\right) \label{BI5}
\end{equation}
is a positive-definite quantity. Hence the
Born-Infeld Lagrangian defined in Eqn.(\ref{BI1})
cannot accelerate the universe (see however \cite{portuga1}).

\subsection*{Appendix III: Static and spherically symmetric
electromagnetic solution and the asymptotic regime}

\begin{figure}[h]
\begin{center}
\includegraphics[angle=-90,width=0.5\textwidth]{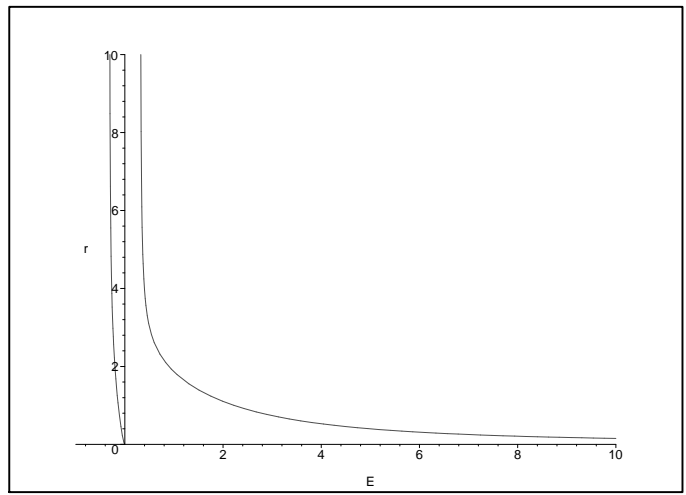}
\caption{Plot of $r=r(E)$ for a positive charge. The plot for negative $q$ is the
mirror image of this plot.}
\label{e1}
\end{center}
\end{figure}

%We have made an analysis of the modification of electrodynamics in a
%cosmological context. We are not arguing that these effects are more
%than the response of the universe to local electrodynamics
%properties. Some decades ago Wheeler and Feynmann made a conjecture
%that local properties of electrodynamics (e.g. the Lienard-Wiechard
%potential) may just be a consequence of such cosmical response
%inducing the elimination of advanced fields....
In this article we have presented a modification of Maxwell's electromagnetism that
can describe in a unified way several phases of the evolution of the universe.
The least we can ask of this modification is that it agrees with
conventional electromagnetism when
%takes these modifications as local change of electrodynamics, we
%should check consistency of the theory with conventional
%electromagnetism. We shall restrict here to the case of the static
the electric field generated by a point charged particle is considered.
For a general
nonlinear Lagrangian $L=L(F)$, Maxwell's equation for a point
charge reduces to
\begin{equation}
r^2 L_F\;E(r)= {\rm const.}
\label{21junho6}
\end{equation}
In the
case of the Lagrangian given in Eqn.(\ref{lag}) we get
\beq
E(r)\left( 4\alpha^2 E(r)^2 - \frac{\gamma}{4E(r) ^2} + \frac 1 4
\right) = \frac{q}{r^2}. \label{ef} \eeq
The polynomial in $E$ that
follows from this equation cannot be solved exactly, but to study
the dependence of $E$ with $r$ we can plot from Eqn.(\ref{ef}) the
function $r=r(E)$ (see Fig.\ref{e1}).
Although the plot displays two branches both for positive
and negative $q$, the
unphysical branches can be discarded without further consequences.
%\begin{figure}[h]
%\begin{center}
%\includegraphics[angle=-90,width=0.5\textwidth]{electric2}
%\caption{.}
%\label{e1}
%\end{center}
%\end{figure}
It also follows from the plot
that $E$ goes to a constant value for $r\rightarrow\infty$. By
taking derivatives of Eqn.(\ref{ef}) it can be shown that the
function $E(r)$ has no extrema \footnote{Note that $E=0$ is not a
solution of Eqn.(\ref{ef}).}. Hence, the modulus of the electric
field decreases monotonically with increasing $r$, from an infinite
value at the origin to a constant (nonzero but small) value at
infinity. Eqn.(\ref{ef}) then shows that $E_\infty = \pm \sqrt{\gamma}$.
This situation is akin to that in the theory defined by the action
$$
S =\frac{M_{\rm P}^2}{2} \int \sqrt{-g} \left( R -
\frac{\alpha^4}{ R} \right) d^4x.
$$
It was shown in \cite{carroll1} that the static and spherically
symmetric solution of this theory does not approach Minkowski
asymptotically: it tends instead to (anti)-de Sitter spacetime.
%Moreover, the electric field of a
%point particle in the nonlinear theory given by
%Eqn.(\ref{lag}) can be made
%
%) .....analogously to the gravitational field in \cite{carroll1}, by
%choosing the parameter $\gamma$ small enough

Regarding the behaviour of the field for small values of $r$,
if we compare the term corresponding to Maxwell's case in Eqn.(\ref{ef})
with the other two, we get that for the field to be Maxwell-like we need
that
$$ \sqrt\gamma << E^2 << \frac{1}{16\alpha^2}.
$$
With the explicit dependence for the field given by
$E(r)=q/r^2$, it would be possible to set a value for
$\alpha$ in agreement with the observation
by
$$
\alpha^2<< \frac{r_0^4}{16q^2},
$$
where $r_0$ is a reference value set by the experiment.

Let us finish this appendix with two comments.
For very weak fields we can neglect the quadratic term in the
Lagrangian, and the energy-momentum tensor reduces to the form
\begin{equation}
T_{\mu\nu} = - L \eta_{\mu\nu} - 4 L_{F} \, F_{\mu\alpha} \,
F^{\alpha}_{{}{\nu}}, \label{21junho7}
\end{equation}
which in the case of a point charge ($F_{01} = E(r)$) is
\begin{eqnarray}
T^{0}_{0}= T^{1}_{1} &=& \frac{1}{2E^{2}} \, \left( E^{4} - 3\mu^{8}
\right),\nonumber \\
 T^{2}_{2}= T^{3}_{3} &=& - \, \frac{1}{2E^{2}} \,
\left( E^{4} + \mu^{8} \right). \label{21junho10}
\end{eqnarray}
In the asymptotic regime we can set $ E \approx \mu^{2}$, obtaining
for the
energy-momentum tensor
\begin{equation}
T^{0}_{0}= T^{1}_{1} =  T^{2}_{2}= T^{3}_{3} = - \, \mu^{4},
\label{21junho11}
\end{equation}
which has the form of a cosmological constant.
By adding an extra term in the Lagrangian (as was done in the case of Born-Infeld)
we could
eliminate the residual constant field at infinity. Such ambiguity does not arise
in the case of
Maxwell's  Electrodynamics due
to the linearity of the equations.
However, for nonlinear electromagnetic theories,
the geometrical structure at
infinity is a matter of choice:
the field is a constant that can be different from zero.
Such a property can be translated in a formal question:
what is the asymptotic regime of the geometry of
space-time: Minkowski or de Sitter?
In linear electrodynamics the answer to that question
is unique, but if nonlinearities are present,
the possibility of a de Sitter structure must
be considered. In theories in which a solution different from zero
for the equation $L_{F} = 0$ exists, such a question has to be
investigated combined with cosmology \footnote{In a recent paper
\cite{none} a new look into this question was considered by
the exam of a proposed relation of the apparent mass of the
graviton and the cosmological constant.}.

\section*{Acknowledgements}
EG acknowledges support of CNPq.
JS is supported by CNPq.
MN acknowledges support of FAPERJ and CNPq.
SEPB acknowledges support from UERJ.
%\begin{figure}[h]
%\begin{center}
%\includegraphics[angle=-90,width=0.5\textwidth]{f1}
%\caption{caption.}
%\label{potencial}
%\end{center}
%\end{figure}
%
%\begin{figure}[h]
%\begin{center}
%\includegraphics[angle=-90,width=0.5\textwidth]{f2}
%\caption{caption.}
%\label{potencial}
%\end{center}
%\end{figure}
%\begin{figure}[h]
%\begin{center}
%\includegraphics[angle=-90,width=0.5\textwidth]{f3}
%\caption{caption.}
%\label{potencial}
%\end{center}
%\end{figure}


\begin{thebibliography}{99}
\bibitem{gamma} Notice that on dimensional grounds,
$ \gamma = \hbar ^2\times \mu^{8}, $ where $\mu$ is a fundamental
constant with dimension of length$^{-1}$, which should be
constrained by observation.
\end{thebibliography}
\end{document}